\documentclass[aps,prl,preprint,groupedaddress]{revtex4}

\usepackage[dvips]{graphicx,color} 
\usepackage{array,hhline,dcolumn} 
\usepackage{rotating} 

\newcommand{\gtwid}{\mathrel{\raise.3ex\hbox{$>$\kern-.75em\lower1ex\hbox{$\sim$}}}}
\newcommand{\ltwid}{\mathrel{\raise.3ex\hbox{$<$\kern-.75em\lower1ex\hbox{$\sim$}}}}

\begin{document}
%

\title{Unexplained Excess of Electron-Like Events From a 1-GeV
Neutrino Beam}

\author{A.~A. Aguilar-Arevalo$^{5}$, C.~E.~Anderson$^{18}$,
        A.~O.~Bazarko$^{15}$, S.~J.~Brice$^{7}$, B.~C.~Brown$^{7}$,
        L.~Bugel$^{5}$, J.~Cao$^{14}$, L.~Coney$^{5}$,
        J.~M.~Conrad$^{5,13}$, D.~C.~Cox$^{10}$, A.~Curioni$^{18}$,
        Z.~Djurcic$^{5}$, D.~A.~Finley$^{7}$, B.~T.~Fleming$^{18}$,
        R.~Ford$^{7}$, F.~G.~Garcia$^{7}$,
        G.~T.~Garvey$^{11}$, C.~Green$^{7,11}$, J.~A.~Green$^{10,11}$,
        T.~L.~Hart$^{4}$, E.~Hawker$^{3,11}$,
        R.~Imlay$^{12}$, R.~A. ~Johnson$^{3}$, G.~Karagiorgi$^{5,13}$,
        P.~Kasper$^{7}$, T.~Katori$^{10}$, T.~Kobilarcik$^{7}$,
        I.~Kourbanis$^{7}$, S.~Koutsoliotas$^{2}$, E.~M.~Laird$^{15}$,
        S.~K.~Linden$^{18}$,J.~M.~Link$^{17}$, Y.~Liu$^{14}$,
        Y.~Liu$^{1}$, W.~C.~Louis$^{11}$,
        K.~B.~M.~Mahn$^{5}$, W.~Marsh$^{7}$, G.~McGregor$^{11}$,
        W.~Metcalf$^{12}$, P.~D.~Meyers$^{15}$,
        F.~Mills$^{7}$, G.~B.~Mills$^{11}$,
        J.~Monroe$^{5,13}$, C.~D.~Moore$^{7}$, R.~H.~Nelson$^{4}$,
        V.~T.~Nguyen$^{5,13}$, P.~Nienaber$^{16}$, J.~A.~Nowak$^{12}$,
        S.~Ouedraogo$^{12}$, R.~B.~Patterson$^{15}$,
        D.~Perevalov$^{1}$, C.~C.~Polly$^{9,10}$, E.~Prebys$^{7}$,
        J.~L.~Raaf$^{3}$, H.~Ray$^{8,11}$, B.~P.~Roe$^{14}$,
        A.~D.~Russell$^{7}$, V.~Sandberg$^{11}$, R.~Schirato$^{11}$,
        D.~Schmitz$^{5}$, M.~H.~Shaevitz$^{5}$, F.~C.~Shoemaker$^{15}$,
        D.~Smith$^{6}$, M.~Soderberg$^{18}$,
        M.~Sorel$^{5}$\footnote{Present Address: IFIC, Universidad de Valencia
	and CSIC, Valencia 46071, Spain},
        P.~Spentzouris$^{7}$, I.~Stancu$^{1}$,
        R.~J.~Stefanski$^{7}$, 
        M.~Sung$^{12}$, H.~A.~Tanaka$^{15}$,
        R.~Tayloe$^{10}$, M.~Tzanov$^{4}$,
        R.~Van~de~Water$^{11}$, 
	M.~O.~Wascko$^{12}$\footnote{Present address: Imperial College 
	London, London, UK},
	D.~H.~White$^{11}$,
        M.~J.~Wilking$^{4}$, H.~J.~Yang$^{14}$,
        G.~P.~Zeller$^{5,11}$, E.~D.~Zimmerman$^{4}$ \\
\smallskip
(The MiniBooNE Collaboration)
\smallskip
}
\smallskip
\smallskip
\affiliation{
$^1$University of Alabama; Tuscaloosa, AL 35487 \\
$^2$Bucknell University; Lewisburg, PA 17837 \\
$^3$University of Cincinnati; Cincinnati, OH 45221\\
$^4$University of Colorado; Boulder, CO 80309 \\
$^5$Columbia University; New York, NY 10027 \\
$^6$Embry Riddle Aeronautical University; Prescott, AZ 86301 \\
$^7$Fermi National Accelerator Laboratory; Batavia, IL 60510 \\
$^8$University of Florida; Gainesville, FL 32611 \\
$^9$University of Illinois; Urbana, IL 61801 \\
$^{10}$Indiana University; Bloomington, IN 47405 \\
$^{11}$Los Alamos National Laboratory; Los Alamos, NM 87545 \\
$^{12}$Louisiana State University; Baton Rouge, LA 70803 \\
$^{13}$Massachusetts Institute of Technology; Cambridge, MA 02139 \\
$^{14}$University of Michigan; Ann Arbor, MI 48109 \\
$^{15}$Princeton University; Princeton, NJ 08544 \\
$^{16}$Saint Mary's University of Minnesota; Winona, MN 55987 \\
$^{17}$Virginia Polytechnic Institute \& State University; Blacksburg, VA
24061
\\
$^{18}$Yale University; New Haven, CT 06520\\
}

\date{\today}

\begin{abstract}
The MiniBooNE Collaboration observes unexplained electron-like events 
in the reconstructed neutrino energy range from 200 to 475 MeV.
With $6.46 \times 10^{20}$ protons on target, 544 electron-like
events are observed in this energy range,
compared to an expectation of $415.2 \pm 43.4$ events, corresponding
to an excess of $128.8 \pm 20.4 \pm 38.3$ events.
The shape of the excess in several kinematic variables is
consistent with being due to either
$\nu_e$ and $\bar \nu_e$
charged-current scattering or to $\nu_\mu$
neutral-current scattering with a photon in the final state. 
No significant excess of events is observed 
in the reconstructed neutrino energy range from 475 to 1250 MeV, where
408 events are observed compared to an expectation of $385.9 \pm 35.7$ events.

\end{abstract}

\pacs{14.60.Lm, 14.60.Pq, 14.60.St}

\keywords{Suggested keywords}
\maketitle


In a previous Letter \cite{mb_osc}, the MiniBooNE collaboration reported
initial results on a search for $\nu_\mu \rightarrow \nu_e$ oscillations.
The search was motivated by the LSND observation \cite{lsnd} of an
excess of $\bar \nu_e$ events in a $\bar \nu_\mu$ beam that implied larger
values of $\Delta m^2$ than any of the currently confirmed oscillation
measurements. The MiniBooNE result showed no evidence of an excess of 
electron-like events for neutrino energies above 475 MeV. However, a 
sizeable excess of electron-like events was observed from 300-475 MeV.
This Letter reports on a more detailed investigation of the low-energy
electron-like events \cite{BDT}. 
Published explanations for the low-energy excess range 
from anomaly mediated neutrino-photon coupling \cite{hhh} 
to neutrino oscillations involving sterile neutrinos 
\cite{sorel,weiler,goldman,maltoni,nelson} to Lorentz violation 
\cite{kostelecky}. In the course of this investigation,
many improvements have been made to the data analysis, and the
data sample has increased from $5.58 \times 10^{20}$ protons on target
(POT) to $6.46 \times
10^{20}$ POT. The excess of electron-like events 
persists after these improvements and has been studied as a
function of several kinematic variables.

MiniBooNE uses the Fermilab Booster
neutrino beam, which is generated from 8-GeV kinetic energy protons
incident on a beryllium production target. Neutrinos are produced
in a 50 m long decay pipe by the in-flight decay of pions and kaons and a
small fraction of the subsequent muons. 
The center of the MiniBooNE detector is 541~m from the production target 
\cite{mb_detector}. The neutrino target and 
detector medium is mineral oil in which
relativistic particles create both Cherenkov and scintillation
light. The different properties of these sources of light readily allow
particle identification; however, the detector cannot distinguish between
electrons and photons.

The Booster neutrino beam flux at the detector is
modeled using a GEANT4-based simulation \cite{geant4} of the
beamline. Pion and kaon production in the target is
parametrized \cite{SW} by a global fit to proton-beryllium
particle production data \cite{harp,prod_data}.
The $\nu_\mu$ energy spectrum peaks at $\sim 600$ MeV, has a mean
energy of $\sim 800$ MeV, and extends 
to $\sim 3000$ MeV \cite{mb_flux}. 

The specific changes to the analysis of the low-energy events since the
initial paper \cite{mb_osc} are discussed in
some detail in the following text. 

The v3 NUANCE \cite{nuance} event generator
is used to simulate neutrino interactions in mineral oil. 
The constraint on neutral-current (NC) 
$\pi^0$ production from MiniBooNE data was expanded 
to finer momentum bins \cite{mb_pi0}.
Also, 
a direct measurement of low energy NC coherent $\pi^0$ production 
\cite{mb_pi0} was
implemented to improve
the modeling of $\pi^0$ events in the most forward direction.
In addition, there is a more accurate treatment
of the ratio of $\gamma$ to $\pi^0$ decay of $\Delta$ in nuclei.
To avoid uncertainties in neutrino flux and NC
cross sections, the number of
$\Delta$ radiative decays is determined from the
number of measured NC $\pi^0$ events.

Final state particles from the initial neutrino
interaction \cite{nuance}, their decays, and possible strong and electromagnetic
re-interactions in the detector
medium are modeled using a GEANT3-based \cite{geant3} simulation, with strong interactions
simulated using
GCALOR \cite{gcalor}. 
Since the previous Letter \cite{mb_osc}, a number of
processes, missing from the
strong interaction model, have been added that could create electron-like
backgrounds:
photonuclear interactions on carbon, radiative $\pi^-$ capture, radiative decay
of $\Delta$ resonances produced in pion-carbon interactions, and $\pi^\pm$-C
(strong) elastic scattering.
Radiative capture and $\Delta \rightarrow N\gamma$
decay produce single photons that MiniBooNE cannot
distinguish from electrons.
Photonuclear interactions
can cause a photon from a $\pi^0$ to be missed, leaving a single photon.
Elastic scattering of charged pions can cause Cherenkov rings 
to appear more electron-like. 
Of these, only photonuclear interactions contribute significantly
to the electron-like background with apparent neutrino energy $>200$ MeV.
The well-measured photonuclear cross section on carbon is used to simulate
final states from excitation of the giant dipole resonance and $\Delta$
production above and below the pion threshold.
The addition of photonuclear absorption increases the estimated
background from NC $\pi^0$ scattering by $\sim30\%$ in the
energy range $200<E_\nu^{QE}<475$ MeV.
$E_\nu^{QE}$
is the reconstructed incident neutrino energy and is determined from
the reconstructed lepton energy and angle with respect
to the known neutrino direction, assuming charged-current quasi-elastic
(CCQE) scattering.

One of the larger $\nu_\mathrm{e}$ backgrounds at low energy results from
neutrino interactions in the 
tank wall and concrete vault and dirt surrounding
the detector. These events
originating outside the detector are uniquely characterized by low visible energy
($E_{vis}$), large radius, 
and a direction that points into the detector; therefore, their contribution can
be measured from MiniBooNE
data. An improved estimate of this background using reconstructed event position
and direction information 
reduces the normalization of such backgrounds by $30\%$. In addition, a new
selection criterion based on 
energy and topology rejects $83\%$ of these events, while discarding only $21\%$
of signal events in the 
$200<E_\nu^{QE}<475$ MeV energy range.

Numerous improvements have been incorporated
in the systematic
error determination associated with the neutrino flux, detector, and neutrino
cross section modeling. In
estimating neutrino flux uncertainties, the propagation of $\pi^+$ production
errors has been upgraded 
to remove unnecessary model dependence. This results in 
a decrease in the overall
$\pi^+$ production 
uncertainty from $\sim16\%$ to $\sim7\%$ \cite{mb_flux}, 
which better reflects the
uncertainties in the
underlying HARP measurement of $\pi^+$ production on Be \cite{harp}.
In the detector simulation, a
comprehensive set of final state variations has been evaluated to conservatively
encompass the uncertainty 
in the aforementioned list of added hadronic processes. These uncertainties
contribute an additional
$1\%$ uncertainty in the low energy MiniBooNE oscillation analysis. In the
neutrino cross section model,
the estimation of the $\Delta$ radiative decays uncertainty has increased from 
$9\%$
to $12\%$. Also,
measurements of the rates of coherently and resonantly produced $\pi^0$ events 
\cite{mb_pi0} has enabled some reduction in these errors.

The reconstruction and selection of electron-like events is identical
to the initial analysis \cite{mb_osc} 
with the addition of the cut to reject
events produced outside the detector described earlier.
Events are reconstructed under
four hypotheses: a single electron-like Cherenkov ring, a single
muon-like ring, two photon-like rings with unconstrained kinematics, and
two photon-like rings with an invariant mass $M_{\gamma\gamma}=m_{\pi^0}$.
To select $\nu_e$-candidate events, an
initial selection is first applied followed by particle identification
cuts. 

Four different analyses are performed on the data.
\begin{itemize}
\item Original Analysis: original analysis \cite{mb_osc} with the
original data set of $5.58 \times 10^{20}$ POT.
\item Revised Analysis: the Original Analysis with the updated background and 
uncertainty estimates described in this paper.
\item Extended Analysis: the Revised Analysis but with the extended data set 
of $6.46 \times 10^{20}$ POT. 
\item Final Analysis: the Extended Analysis but including the new 
external event cut.
\end{itemize}

Table \ref{signal_bkgd} shows the 
expected number of events with 
$E_\nu^{QE}$ between $200 - 300$ MeV, $300 - 475$ MeV, and
$475 - 1250$ MeV after
the complete event selection of the Final Analysis.
The background estimates include
antineutrino events, representing $<2\%$ of the total.
The total expected backgrounds for the three energy regions are 
$186.8 \pm 26.0$ events, $228.3 \pm 24.5$ events, 
and $385.9 \pm 35.7$ events, respectively.

\begin{table}
\caption{\label{signal_bkgd} \em The expected number of events
in the $200<E_\nu^{QE}<300$ MeV, $300<E_\nu^{QE}<475$ MeV,
and $475<E_\nu^{QE}<1250$ MeV 
energy ranges from all of the backgrounds
after the complete event selection of the Final Analysis.}
\begin{ruledtabular}
\begin{tabular}{cccc}
Process&$200-300$&$300-475$&$475-1250$ \\
\hline
$\nu_\mu$ CCQE&9.0&17.4&11.7 \\
$\nu_\mu e \rightarrow \nu_\mu e$&6.1&4.3&6.4 \\
NC $\pi^0$&103.5&77.8&71.2 \\
NC $\Delta \rightarrow N \gamma$&19.5&47.5&19.4 \\
External Events&11.5&12.3&11.5 \\
Other Events&18.4&7.3&16.8 \\
\hline
$\nu_e$ from $\mu$ Decay&13.6&44.5&153.5 \\
$\nu_e$ from $K^+$ Decay&3.6&13.8&81.9 \\
$\nu_e$ from $K^0_L$ Decay&1.6&3.4&13.5 \\
\hline
Total Background &$186.8 \pm 26.0$&$228.3 \pm 24.5$&$385.9 \pm 35.7$ \\
\end{tabular}
\end{ruledtabular}
\end{table}

A total of 
1069 events pass the complete event selection of the Final Analysis
with $E_\nu^{QE}>200$ MeV.
The numbers of data, background, and excess events for different
$E_\nu^{QE}$ ranges are shown in Table \ref{signal_bkgd3},
together with the significance of the excesses for the four analyses.
The uncertainties include both statistical and systematic errors.
While
there is no significant event excess for $E_\nu^{QE}>475$ MeV,
a sizeable excess is observed for $E_\nu^{QE}<475$ MeV.
For the Final Analysis,
an excess of $128.8 \pm 20.4 \pm 38.3$ events ($3.0 \sigma$)
is observed for $200<E_\nu^{QE}<475$ MeV. 

\begin{figure}[htbp]
\centerline{\includegraphics[height=2.0in]{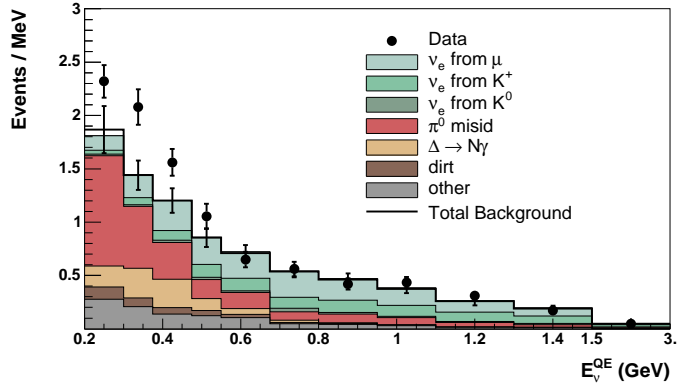}}
\caption{The $E_\nu^{QE}$ distribution for data (points with 
statistical errors) and backgrounds (histogram with systematic errors).}
\label{data_mc1}
\end{figure}

\begin{figure}[htbp]
\centerline{\includegraphics[height=1.9in]{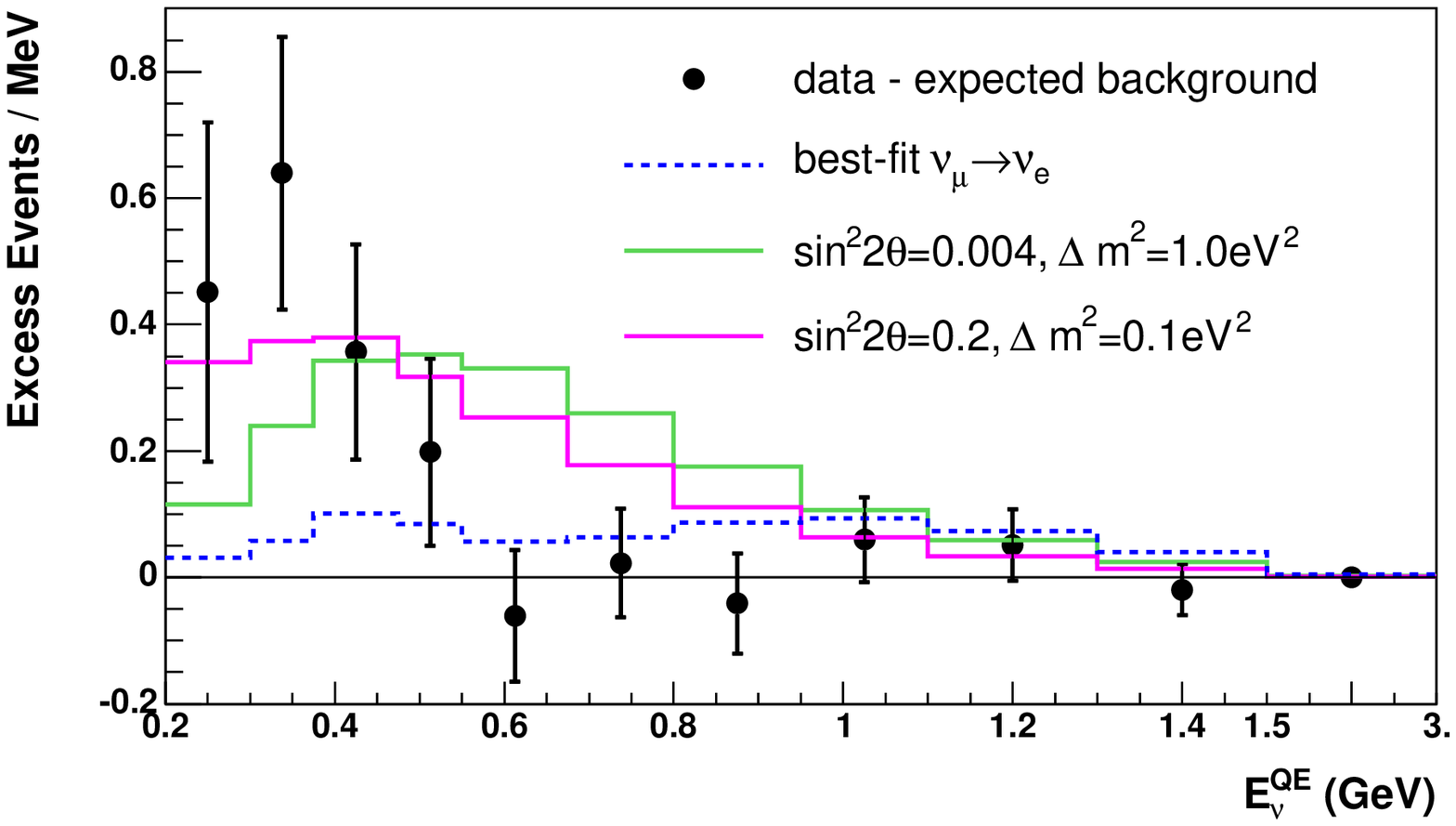}}
\caption{The event excess as a function of $E_\nu^{QE}$. Also shown are
the expectations from 
the best oscillation fit and from 
neutrino oscillation parameters in the LSND
allowed region \cite{lsnd}. The error bars include both statistical and systematic errors.}
\label{data_mc2}
\end{figure}

\begin{figure}[htbp]
\centerline{\includegraphics[height=3.3in,angle=-90]{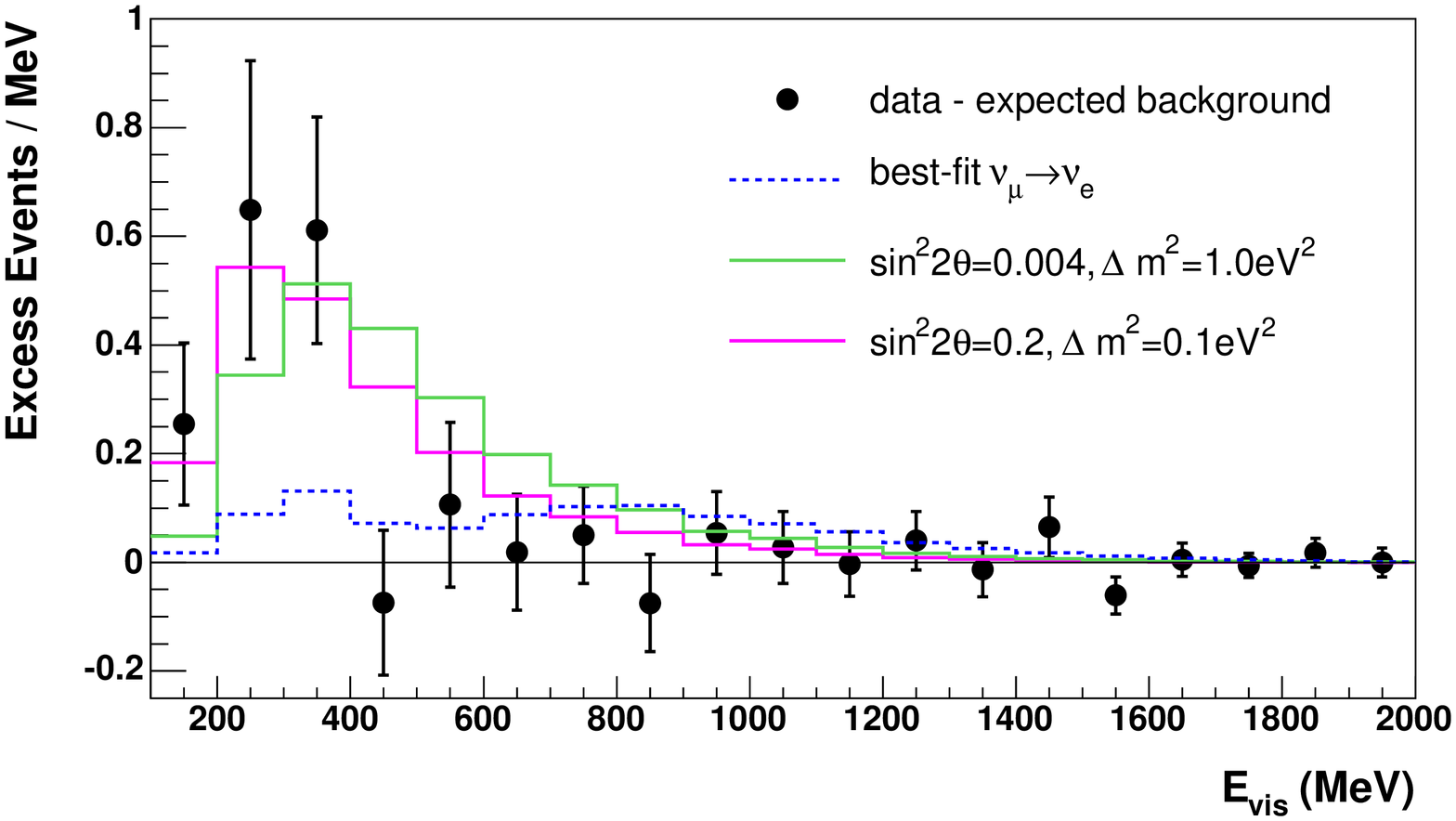}}
\caption{The event excess as a
function of $E_{vis}$ for $E_\nu^{QE} > 200$ MeV. Also shown are
the expectations from 
the best oscillation fit and from
neutrino oscillation parameters in the LSND
allowed region \cite{lsnd}. The error bars include both statistical and systematic errors.}
\label{data_mc3}
\end{figure}

Figure \ref{data_mc1}
shows the $E_\nu^{QE}$ distribution for data (points with 
statistical errors) and backgrounds (histogram with systematic
errors) for the Final Analysis, and
Fig.~\ref{data_mc2} shows the event excess as a function of 
$E_\nu^{QE}$. Also shown in the figure, for comparison, are
expectations from the 
best oscillation fit and from
neutrino oscillation parameters in the LSND
allowed region \cite{lsnd}, which are ruled out at 95\% CL if
the data are fit with $E_\nu^{QE}>475$ MeV \cite{mb_osc}.
The error bars include both statistical and systematic errors.
The best ocillation fit for $E_\nu^{QE}>200$ MeV 
corresponds to $\Delta m^2 = 3.14$ eV$^2$
and $\sin^22\theta = 0.0017$ and has a $\chi^2/DF=18.3/17$. The null
fit has a $\chi^2/DF=22.0/19$. For $E_\nu^{QE}>475$ MeV, the best fit
is consistent with the initial result of no oscillations \cite{mb_osc}.
As shown in Fig.~\ref{data_mc3} for $E_\nu^{QE} >200$ MeV,
the event excess 
occurs for $E_{vis} <400$ MeV.

\begin{figure}[htbp]
\centerline{\includegraphics[height=1.95in]{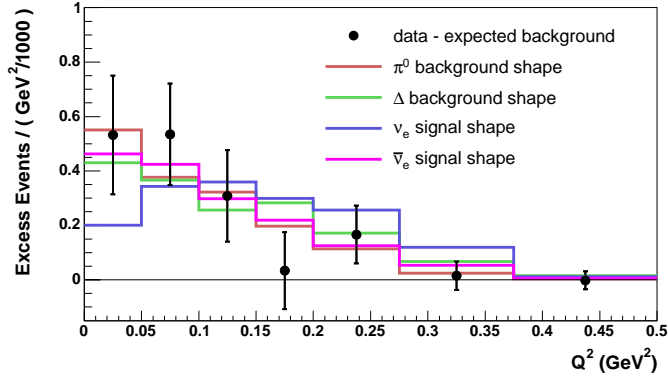}}
\caption{The event excess as a
function of $Q^2$ for $300 < E_\nu^{QE} < 475$ MeV. Also shown 
are the
expected shapes from the NC $\pi^0$ and $\Delta \rightarrow N \gamma$
reactions, which are representative of photon events
produced by NC scattering, and from CC
$\nu_e C \rightarrow e^- X$ and
$\bar \nu_e C \rightarrow e^+ X$ scattering.
The error bars include both
data statistical and shape-only systematic errors.}
\label{data_mc4}
\end{figure}

\begin{figure}[htbp]
\centerline{\includegraphics[height=1.95in]{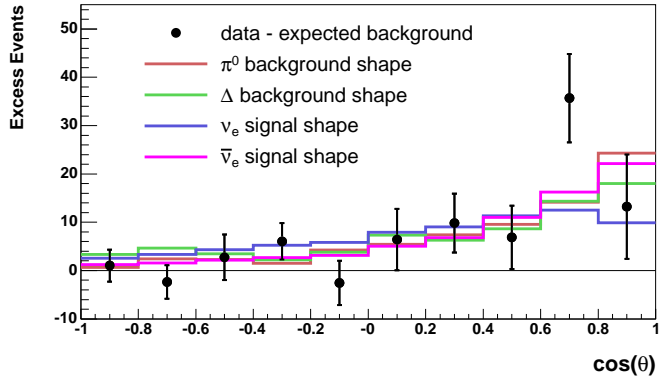}}
\caption{The event excess as a
function of $\cos (\theta)$ for $300 < E_\nu^{QE} < 475$ MeV. 
The legend is the same as Fig. \ref{data_mc4}.}
\label{data_mc5}
\end{figure}

Figs.~\ref{data_mc4} and \ref{data_mc5} show the event excess as  
functions of reconstructed 
$Q^2$ and $\cos (\theta)$ for $300 < E_\nu^{QE} < 475$ MeV, the
energy region with the most significant excess. 
$Q^2$ is determined from the energy and angle of the outgoing lepton,
assuming CCQE scattering,
and $\theta$ is the angle between the incident neutrino and outgoing lepton.
Also shown in the figures are the
expected shapes from 
the NC $\pi^0$ and $\Delta \rightarrow N \gamma$
reactions, which are representative of photon events
produced by NC scattering, and from $\nu_e C \rightarrow e^- X$ and
$\bar \nu_e C \rightarrow e^+ X$ CC scattering. 
The different reactions all assume the same $\nu_\mu$ energy spectrum.
As shown in Table \ref{chisquare}, the $\chi^2$ values from
comparisons of the event excess to the expected shapes are acceptable for all
of the processes. Also shown in the table is the factor increase necessary
for each process
to explain the low-energy excess. In each case,
the estimated background would have to more than double 
(increase  by $>5 \sigma$) to explain the excess.

\begin{table*}
\caption{\label{signal_bkgd3} \em The number of data, 
background, and excess events for different
$E_\nu^{QE}$ ranges, together with the
significance of the excesses. The different analyses
are described in the text.}
\begin{ruledtabular}
\begin{tabular}{cccc|c}
Event Sample&Original Analysis \cite{mb_osc}&Revised Analysis&Extended Analysis&
{\bf Final Analysis} \\
\hline
$200-300$ MeV&&&& \\
Data&375&368&427&232 \\
Background&$283 \pm 37$&$332.4 \pm 38.9$&$386.0 \pm 44.3$&$186.8 \pm 26.0$ \\
Excess (Significance)&$92 \pm 37$ ($2.5 \sigma$)&$35.6 \pm 38.9$ ($0.9 \sigma$)&$41.0 \pm 44.3$ ($0.9 \sigma$)&$45.2 \pm 26.0$ ($1.7 \sigma$) \\
\hline
$300-475$ MeV&&&& \\
Data&369&364&428&312 \\
Background&$273 \pm 26$&$282.9 \pm 28.3$&$330.0 \pm 31.8$&$228.3 \pm 24.5$ \\
Excess (Significance)&$96 \pm 26$ ($3.7 \sigma$)&$81.1 \pm 28.3$ ($2.9 \sigma$)&$98.0 \pm 31.8$ ($3.1 \sigma$)&$83.7 \pm 24.5$ ($3.4 \sigma$) \\
\hline
$200-475$ MeV&&&& \\
Data&744&732&855&544 \\
Background&$556 \pm 54$&$615.3 \pm 58.0$&$716.1 \pm 66.2$&$415.2 \pm 43.4$ \\
Excess (Significance)&$188 \pm 54$ ($3.5 \sigma$)&$116.7 \pm 58.0$ ($2.0 \sigma$)&$138.9 \pm 66.2$ ($2.1 \sigma$)&$128.8 \pm 43.4$ ($3.0 \sigma$) \\
\hline
$475-1250$ MeV&&&& \\
Data&380&369&431&408 \\
Background&$358 \pm 40$&$356.0 \pm 33.3$&$412.7 \pm 37.6$&$385.9 \pm 35.7$ \\
Excess (Significance)&$22 \pm 40$ ($0.6 \sigma$)&$13.0 \pm 33.3$ ($0.4 \sigma$)&$18.3 \pm 37.6$ ($0.5 \sigma$)&$22.1 \pm 35.7$ ($0.6 \sigma$) \\
\end{tabular}
\end{ruledtabular}
\end{table*}

\begin{table}
\caption{\label{chisquare} \em The 
$\chi^2$ values 
from comparisons of the event excess $Q^2$ and $\cos (\theta)$ distributions
for $300 < E_\nu^{QE} < 475$ MeV 
to the expected shapes from 
various NC and CC reactions.
Also shown is the factor increase necessary 
for the estimated background for each process
to explain the low-energy excess.}
\begin{ruledtabular}
\begin{tabular}{cccc}
Process&$\chi^2(cos \theta)/9$ DF&$\chi^2(Q^2)/6$ DF&Factor Increase \\
\hline
NC $\pi^0$&13.46&2.18&2.0 \\
$\Delta \rightarrow N \gamma$&16.85&4.46&2.7 \\
$\nu_e C \rightarrow e^- X$&14.58&8.72&2.4 \\
$\bar \nu_e C \rightarrow e^+ X$&10.11&2.44&65.4 \\
\end{tabular}
\end{ruledtabular}
\end{table}

In summary, MiniBooNE observes an unexplained excess of 
$128.8 \pm 20.4 \pm 38.3$ electron-like events in the
energy region $200< E_\nu^{QE} < 475$ MeV. These events are consistent with
being either electron events produced by CC
scattering ($\nu_e C \rightarrow e^- X$ or $\bar \nu_e C \rightarrow e^+ X$) 
or photon events 
produced by NC 
scattering ($\nu C \rightarrow \nu \gamma X$). 
Upcoming MiniBooNE results
with the Booster antineutrino beam and with the NuMI neutrino beam 
\cite{mb_numi} 
should help distinguish these two possibilities and shed 
further light on the
low-energy region.

\bigskip

\begin{acknowledgments}
We acknowledge the support of Fermilab, the Department of Energy,
and the National Science Foundation, and
we acknowledge Los
Alamos National Laboratory for LDRD funding. In addition, we 
acknowledge theoretical input from Tina Leitner and Ulrich 
Mosel on the $\Delta \rightarrow N \gamma$ background.
\end{acknowledgments}


\newpage 
\bibliography{prl}

\end{document}